

\font\twelverm=cmr10 scaled 1200    \font\twelvei=cmmi10 scaled 1200
\font\twelvesy=cmsy10 scaled 1200   \font\twelveex=cmex10 scaled 1200
\font\twelvebf=cmbx10 scaled 1200   \font\twelvesl=cmsl10 scaled 1200
\font\twelvett=cmtt10 scaled 1200   \font\twelveit=cmti10 scaled 1200

\skewchar\twelvei='177   \skewchar\twelvesy='60


\def\twelvepoint{\normalbaselineskip=12.4pt
  \abovedisplayskip 12.4pt plus 3pt minus 9pt
  \belowdisplayskip 12.4pt plus 3pt minus 9pt
  \abovedisplayshortskip 0pt plus 3pt
  \belowdisplayshortskip 7.2pt plus 3pt minus 4pt
  \smallskipamount=3.6pt plus1.2pt minus1.2pt
  \medskipamount=7.2pt plus2.4pt minus2.4pt
  \bigskipamount=14.4pt plus4.8pt minus4.8pt
  \def\rm{\fam0\twelverm}          \def\it{\fam\itfam\twelveit}%
  \def\sl{\fam\slfam\twelvesl}     \def\bf{\fam\bffam\twelvebf}%
  \def\mit{\fam 1}                 \def\cal{\fam 2}%
  \def\tt{\twelvett}
  \def\nullspace{\nulldelimiterspace=0pt \mathsurround=0pt }
  \def\big##1{{\hbox{$\left##1\vbox to 10.2pt{}\right.\nullspace$}}}
  \def\Big##1{{\hbox{$\left##1\vbox to 13.8pt{}\right.\nullspace$}}}
  \def\bigg##1{{\hbox{$\left##1\vbox to 17.4pt{}\right.\nullspace$}}}
  \def\Bigg##1{{\hbox{$\left##1\vbox to 21.0pt{}\right.\nullspace$}}}
  \textfont0=\twelverm   \scriptfont0=\tenrm   \scriptscriptfont0=\sevenrm
  \textfont1=\twelvei    \scriptfont1=\teni    \scriptscriptfont1=\seveni
  \textfont2=\twelvesy   \scriptfont2=\tensy   \scriptscriptfont2=\sevensy
  \textfont3=\twelveex   \scriptfont3=\twelveex  \scriptscriptfont3=\twelveex
  \textfont\itfam=\twelveit
  \textfont\slfam=\twelvesl
  \textfont\bffam=\twelvebf \scriptfont\bffam=\tenbf
  \scriptscriptfont\bffam=\sevenbf
  \normalbaselines\rm}



\def\beginlinemode{\endmode
  \begingroup\parskip=0pt \obeylines\def\\{\par}\def\endmode{\par\endgroup}}
\def\beginparmode{\endmode
  \begingroup \def\endmode{\par\endgroup}}
\let\endmode=\par
{\obeylines\gdef\
{}}
\def\singlespace{\baselineskip=\normalbaselineskip}

\def\oneandahalfspace{\baselineskip=\normalbaselineskip
  \multiply\baselineskip by 3 \divide\baselineskip by 2}
\def\doublespace{\baselineskip=\normalbaselineskip \multiply\baselineskip by 2}

\newcount\firstpageno
\firstpageno=2
\footline={\ifnum\pageno<\firstpageno{\hfil}\else{\hfil\twelverm\folio\hfil}\fi}
\let\rawfootnote=\footnote		
\def\footnote#1#2{{\rm\singlespace\parindent=0pt\rawfootnote{#1}{#2}}}
\def\raggedcenter{\leftskip=4em plus 12em \rightskip=\leftskip
  \parindent=0pt \parfillskip=0pt \spaceskip=.3333em \xspaceskip=.5em
  \pretolerance=9999 \tolerance=9999
  \hyphenpenalty=9999 \exhyphenpenalty=9999 }


\hsize=6.5truein
\vsize=8.9truein
\parskip=\medskipamount
\twelvepoint		
\doublespace		
\overfullrule=0pt	


\def\s{\scriptscriptstyle}

\def\title			
  {\null\vskip 3pt plus 0.2fill
   \beginlinemode \doublespace \raggedcenter \bf}

\def\author			
  {\vskip 3pt plus 0.2fill \beginlinemode
   \singlespace \raggedcenter}

\def\affil			
  {\vskip 3pt plus 0.1fill \beginlinemode
   \oneandahalfspace \raggedcenter \sl}

\def\abstract			
  {\vskip 3pt plus 0.3fill \beginparmode
   \doublespace \narrower  }

\def\endtitlepage		
  {\endpage			
   \body}

\def\body			
  {\beginparmode}		

\def\head#1{			
  \filbreak\vskip 0.5truein	
  {\immediate\write16{#1}
   \raggedcenter \uppercase{#1}\par}
   \nobreak\vskip 0.25truein\nobreak}

\def\subhead#1{			
  \vskip 0.25truein		
  {\raggedcenter #1 \par}
   \nobreak\vskip 0.25truein\nobreak}

\def\refto#1{$^{#1}$}		

\def\references			
  {\head{References}		
   \beginparmode
   \frenchspacing \parindent=0pt \leftskip=1truecm
   \parskip=8pt plus 3pt \everypar{\hangindent=\parindent}}

\gdef\refis#1{\indent\hbox to 0pt{\hss#1.~}}	

\gdef\journal#1, #2, #3, 1#4#5#6{		
    {\sl #1~}{\bf #2}, #3, (1#4#5#6)}		

\gdef\journ2 #1, #2, #3, 1#4#5#6{		
    {\sl #1~}{\bf #2}: #3, (1#4#5#6)}		

\def\refstylenp{		
  \gdef\refto##1{ [##1]}				
  \gdef\refis##1{\indent\hbox to 0pt{\hss##1)~}}	
  \gdef\journal##1, ##2, ##3, ##4 {			
     {\sl ##1~}{\bf ##2~}(##3) ##4 }}

\def\refstyleprnp{		
  \gdef\refto##1{ [##1]}				
  \gdef\refis##1{\indent\hbox to 0pt{\hss##1)~}}	
  \gdef\journal##1, ##2, ##3, 1##4##5##6{		
    {\sl ##1~}{\bf ##2~}(1##4##5##6) ##3}}

\def\figurecaptions		
  {\endpage
   \beginparmode
   \head{Figure Captions}
}

\def\endpage			
  {\vfill\eject}

\def\endpaper			
  {\endmode\vfill\supereject}

\def\endit
  {\endpaper\end}


\def\ref#1{Ref. #1}			

\def\frac#1#2{{\textstyle #1 \over \textstyle #2}}

\def\half{{\textstyle {1 \over 2}}}

\def\etal{{\it et al.}}

\def\sla{\raise.15ex\hbox{$/$}\kern-.57em}
\def\leaderfill{\leaders\hbox to 1em{\hss.\hss}\hfill}
\def\twiddle{\lower.9ex\rlap{$\kern-.1em\scriptstyle\sim$}}
\def\bigtwiddle{\lower1.ex\rlap{$\sim$}}
\def\gtwid{\mathrel{\raise.3ex\hbox{$>$\kern-.75em\lower1ex\hbox{$\sim$}}}}
\def\ltwid{\mathrel{\raise.3ex\hbox{$<$\kern-.75em\lower1ex\hbox{$\sim$}}}}
\def\square{\kern1pt\vbox{\hrule height 1.2pt\hbox{\vrule width 1.2pt\hskip 3pt
   \vbox{\vskip 6pt}\hskip 3pt\vrule width 0.6pt}\hrule height 0.6pt}\kern1pt}


\title

Currents in the compressible and incompressible regions
of the two-dimensional electron gas

\vskip 0.2 in

\author

Michael R. Geller and Giovanni Vignale

\vskip 0.01 in

\affil

Department of Physics
University of Missouri
Columbia, Missouri 65211

\abstract

We derive a general expression for the low-temperature
current distribution in a two-dimensional
electron
gas, subjected to a perpendicular magnetic field and in a confining
potential that varies slowly on the scale of the magnetic length
$\ell$.
The analysis is based on a self-consistent one-electron description, such
as the Hartree or standard Kohn-Sham equations.
Our expression, which correctly describes the current distribution on scales
larger than $\ell$, has two components:
One is an ``edge current'' which is proportional to the local
density gradient and the other is a ``bulk current'' which is proportional
to the gradient of the confining potential.
The direction of these currents generally display a striking alternating
pattern.
In a compressible region at the edge of the $n$th Landau level,
the magnitude of the edge current is simply
${\bf j}=-e \omega_{\rm c} \ell^2 (n+\half) \nabla \rho \times {\bf e}_z $
where $\omega_{\rm c}$ is the cyclotron frequency
and $\rho$ is the electron density.
The bulk component, a Hall current,
dominates in the incompressible regions.
In the ideal case of perfect compressibility and incompressibility,
only one type of current contributes to a given region, and
the integrated currents in these regions are universal, independent of the
widths, positions, and geometry of the regions.
The integrated current in the $n$th edge channel is
$(n+\half)e \omega_{\rm c} / 2 \pi$, whereas in an incompressible
strip with integral filling factor $\nu$ it is $\nu e \omega_{\rm c} / 2 \pi$
with
the opposite sign.

\vskip 0.010 in

\noindent PACS numbers: 73.20.Dx, 73.40.--c, 73.40.Hm, 73.50.--h

\endtitlepage

\body

\head{I. INTRODUCTION}

In the past several years, there has been tremendous interest in the
low-temperature properties of a two-dimensional (2D) electron gas
in a strong magnetic field.
The most common experimental realization of this system is at a
modulation-doped semiconductor heterojunction, grown by
molecular beam epitaxy.
One important question, now receiving much
attention, is the distribution of current when the system is subjected to a
 confining
potential, or to an applied voltage, or to both. Knowledge of the equilibrium
 and
nonequilibrium current distributions is important for understanding
the quantum Hall effect, the electronic properties of
low-dimensional semiconductor
nanostructures such as quantum dots or quantum wires in the presence of a
magnetic field, and for understanding mesoscopic transport in general.
The current distribution can also be used to calculate the magnetic properties
(for example, the orbital magnetization) of a confined 2D
electron gas.

Two types of methods are commonly used to obtain a confined
2D electron gas.
Lithographic methods
result in the well-known etched structures consisting of
a patterned region of the
2D electron gas along with its compensating positively charged donors.
The second method produces a confining potential via one or more
evaporated metal gates.
In both cases, the actual confining potential is the sum of the potential from
the remote donors and gates, plus the self-consistent electrostatic
potential (and possibly the exchange-correlation scalar potential) of
the electrons.
Typical electron sheet densities in the modulation-doped 2D electron gas vary
from $10^{11}$ to $10^{12}$ cm$^{-2}$.
The most interesting effects of an applied magnetic field then occur at field
strengths ranging from $1$ to $10$ Tesla, where only a few Landau levels are
occupied.
At these high field strengths, the magnetic length, ranging from
$50$ to a few hundred \AA, is small compared with the length scale over
which the confining potential changes by $\hbar \omega_{\rm c}$,
where $\omega_{\rm c}$ is the cyclotron frequency. In this sense, the
confining potential is {\it slowly varying}.

Vignale and Skudlarski\refto{1} have recently used
current-density functional theory\refto{2}
to derive an exact formal relation between the ground-state current
and density distributions of a three-dimensional
interacting electron gas in the presence of a magnetic field.
In the local density approximation (LDA), valid when the density varies
slowly on the scale of the magnetic length, they obtain
an explicit formula for the current in terms of the density gradient with a
coefficient of proportionality involving thermodynamic quantities
for a uniform electron gas.
However, the application of this LDA result to the 2D electron gas
is complicated by the presence of incompressible regions, where
the density gradient vanishes and the coefficient diverges.
Furthermore, in the 2D electron gas, the density may change from one integral
 filling
factor to another over a single magnetic length, invalidating the {\it a
priori}
application of the LDA.
This has motivated us to reexamine the relation between current and density
in two dimensions, without using the LDA.

Another motivation for this work is to study the
equilibrium currents in edge channels.\refto{3-7}
In a recent paper, Chklovskii \etal\refto{5} have calculated the
classical electrostatic potential and density of a gate-confined
2D electron gas. They show that the electrostatic potential
consists of a series of wide steps of height
$\hbar \omega_{\rm c}$.
In contrast to a naive single-electron picture, there are wide compressible
 regions
where the density gradually changes from one integral filling factor to
another, the so-called edge channels, and narrow incompressible
regions of integral filling factor.
This type of bahavior had been previously noted by McEuen \etal\refto{6}
in the context of quantum dots.
The classical electrostatic analysis of Chklovskii \etal $\ $
has also been extended to narrow gate-confined channels\refto{7} and to
quantum dots.\refto{8} The electrostatics of edge channels in mesa-etched
samples has been studied by Gelfand and Halperin,\refto{9}
and considerable effort has been recently
devoted to extending the classical electrostatic treatment to
self-consistent Hartree and Hartree-Fock approximations.\refto{9-13}
A related question is the transition between sharp and smooth density
 distributions
as the slope of the confining potential is changed.\refto{13,14}

Prior to the recent work on edge channels, considerable progress had already
 been
made in understanding the distribution of current
in the quantum Hall regime.
In one of the early papers on this subject,
MacDonald \etal\refto{15} used the localized nature of the Landau states
to show that the ground-state current density, directed in the
$y$ direction along a Hall bar whose confining potential
$V(x)$ varies in the $x$ direction only, is
simply proportional to $V'(x)$ in the interior of the Hall bar.
Several authors\refto{15-20} have calculated nonequilibrium density and
current distributions in particular confining potentials,
and the correct description of the nonequilibrium steady state
is now a problem of great interest.\refto{18-23}
In particular, Thouless has emphasized the importance of
nonequilibrium bulk currents that are
induced by edge-charge redistribution.\refto{20}

In this paper, we derive the low-temperature
$( k_{\rm \s B} T < < \hbar \omega_{\rm c})$  density and current distributions
 for
a high-mobility spin-polarized 2D electron gas in an arbitrary confining
 potential
$V({\bf r})$ and uniform magnetic field
${\bf B} = B {\bf e}_z$, assuming only that the potential varies slowly
on the scale of the magnetic length $\ell \equiv (\hbar c /e B)^{1 \over 2}$.
As stated above, the confining potential is assumed to consist of an
external potential, from remote donor centers and gates, and a self-consistent
Hartree potential or Hartree plus exchange-correlation scalar potential.
A study of exchange and correlation effects, based on current-density
functional theory,\refto{2} shall be published elsewhere.
We now briefly summarize our results.

We shall show that the low-temperature equilibrium current density (for
 electrons
of charge $-e$) may be written as
$ {\bf j} = {\bf j}_{\rm edge}
+ {\bf j}_{\rm bulk},$
where
$$ {\bf j}_{\rm edge} \equiv
- e \omega_{\rm c} \ell^2  \sum_{n=0}^{\infty} (n + \half)
\ \nabla \rho_n({\bf r}) \times {\bf e}_z                   \eqno(1.1)$$
and
$$ {\bf j}_{\rm bulk} \equiv - {e  \over m \omega_{\rm c}}  \ \rho({\bf r})
\ \nabla V({\bf r})
\times {\bf e}_z   .                            \eqno(1.2)$$
Here $\omega_{\rm c} \equiv e B / m c$ is the cyclotron frequency.
The electron number density, $\rho$, is given by
$$ \rho = {1 \over 2 \pi \ell^2}
\sum_{n=0}^{\infty}\ f\big[\hbar\omega_{\rm c}(n+\half )
+ V({\bf r}) - \mu \big] ,
 \eqno(1.3)$$
where $\mu$ is the chemical potential,
$f(\epsilon) \equiv \big(e^{\epsilon / k_{\rm B} T }+1\big)^{-1}$
is the Fermi distribution function,
and
$ \rho_n $
is simply the $n$th term in (1.3).
At low temperatures, the electron density is uniform everywhere except near a
 number
of edges, where the density changes by an amount $(2 \pi \ell^2)^{-1}$.
These compressible regions, or edge channels, follow lines of
constant confining potential.
The edge current (1.1) is a sum of nonoverlapping parts; the
contribution from the edge of the $n$th Landau level is
${\bf j}_{\rm edge} = - e \omega_{\rm c} \ell^2 (n+\half)
\nabla \rho \times {\bf e}_z$.
In terms of the local electric
field ${\bf E} = \nabla V / e$,
the bulk current (1.2)
may also be written as
$ j^i_{\rm bulk} = \sigma^{ij}({\bf r}) \ E^j({\bf r})$,
where
$\sigma^{ij}({\bf r})$
is the local Hall conductivity tensor.
We shall show that the sign difference in the edge
current relative to the bulk current,
which follows from the fact that
$\nabla \rho$ and $\nabla V$ are antiparallel,
leads to striking oscillations in the
direction of the current in a confined 2D electron gas.

The organization of this paper is as follows.
In the next section, we construct the single-particle
Green's function for the confined electron gas
by a gradient expansion in the confining potential.
We keep terms corresponding to the local magnitude of the confining
potential and its first derivative. The neglect of higher order derivatives
results in distributions which are correct on scales larger than $\ell$.
We then use this Green's function to compute the
equilibrium density
and current distributions.
In Section III, we compare our
results with exact distributions obtained for the case
of noninteracting electrons with parabolic confinement.
In Section IV, we study three applications.
First, we calculate the low-temperature current distribution in a
stepped potential characteristic of the self-consistent Hartree
potential for a narrow gate-confined Hall bar, a system well-known to
possess alternating strips of compressible and incompressible electron gas.
Second, we show that
the integrated current in an ideal incompressible strip with integral filling
factor $\nu$ is $\nu e \omega_{\rm c} / 2 \pi$, independent of
the strip position, geometry, and
 width.
Similarly, the
magnitude of the integrated current in an ideal edge channel at the edge of the
$n$th Landau level is found to be $(n+\half)e \omega_{\rm c}/ 2 \pi$,
independent of the channel position, geometry, and width.
The directions of these currents display an alternating pattern.
As a third application, we show that the total azimuthal current in a
quantum dot, as a function of particle number, is quantized in units
of $e \omega_{\rm c}/ 4 \pi$.
In an appendix we show that our expression for the current density
also follows from the long-wavelength limit of the appropriate
linear response functions, and we also show that a careful treatment of the
divergences
in the LDA relation between current and density also leads to our result.
The physical origin of the alternating signs of the edge and bulk
currents is also explained there.

\head{II. EQUILIBRIUM CURRENT DISTRIBUTION}

We shall first obtain the
single-particle Green's function for the confined electron gas
by the following method: (i) First, a Dyson equation is obtained for the
Green's function $G$ of the confined electron gas, in terms of the
Green's function $G^0$ of the uniform electron gas
and the confining potential $V({\bf r})$; (ii) Then
the short-ranged nature of $G^0$ is used to separate the potential near
${\bf r}$ into a local constant potential,
$V({\bf \bar r})$, and a gradient,
$({\bf r}-{\bf \bar r}) \cdot \nabla V({\bf \bar r})$, for some
${\bf \bar r}$ near ${\bf r}$; (iii) Next, the local constant potential
terms are summed to all orders, resulting in a
Green's function $G^1$; (iv) The gradient terms are then treated to first
order, resulting in a first order gradient expansion for $G$;
(v) Finally, the local
direction of the potential gradient is used to find a gauge
in which the Green's function
takes a particularly simple form, corresponding to
an expansion in eigenstates that are
localized in the $\nabla V({\bf \bar r})$ direction.

As stated above, we consider a 2D electron gas in a uniform magnetic
field ${\bf B} = B {\bf e}_z$ and in a slowly varying potential
$V({\bf r})$, where ${\bf r} = (x,y)$.
We assume that the electrons are spin-polarized by the strong
magnetic field, and we disregard the resulting constant
Zeeman energy. The Hamiltonian
may be written as
$$ H = H^0 + V,                         \eqno(2.1)$$
where
$ H^0 \equiv {1 \over 2m}\big( {\bf p} + {e \over c} {\bf A}\big)^2 $
is the Hamiltonian for an electron in the presence of the magnetic field
alone.
In terms of the exact normalized eigenstates
$\Psi_\alpha$ and eigenvalues $E_\alpha$ of $H$, the Green's function
for the confined electron gas
may be written as
$$ G({\bf r},{\bf r}',s) = \sum_\alpha \ {\Psi_\alpha({\bf r})
\Psi^*_\alpha({\bf r}') \over s - E_\alpha} ,                \eqno(2.2)$$
where $s$ is a complex energy variable.
Knowledge of the Green's function
allows one to determine the equilibrium number density,
$$ \rho({\bf r})
= \oint {ds \over 2\pi i} \ f[s-\mu] \ G({\bf r},{\bf r},s),       \eqno(2.3)$$
and current density,
$$ {\bf j}({\bf r}) = - {e \over m} \oint {ds \over 2 \pi i} \ f[s-\mu] \
 \lim_{{\bf r}'
\rightarrow {\bf r}} \ {\rm Re} \bigg( -i \hbar
\nabla + {e \over c} {\bf A} \bigg) G({\bf r},{\bf r}',s),    \eqno(2.4)$$
at fixed chemical potential $\mu$.
The contour in the complex energy plane is to be taken
in the positive sense around the poles of $G$ on the real $s$ axis,
avoiding the poles of $f$.

The Green's function
$G^0({\bf r},{\bf r}',s)$ for the unconfined electron gas
may be
written as
$$ G^0({\bf r},{\bf r}',s)
= \sum_n \int_{-\infty}^{\infty} dq
\ { \phi_{nq}({\bf r}) \phi_{nq}^*({\bf r}') \over
s - \hbar \omega_{\rm c}(n+{\half}) }           ,       \eqno(2.5)$$
where the $\phi_{nq} \ (n=0,1,2,\cdots)$ are the normalized eigenstates
of $H^0$,
$$ H^0 \ \phi_{nq} = \hbar \omega_{\rm c} (n + \half) \ \phi_{nq}.
\eqno(2.6)$$
In the gauge
$ {\bf A} = B x {\bf e}_y $
these eigenstates are
$$ \phi_{nq} = C_n \ e^{-iqy} \ e^{-{1 \over 2} ({x \over \ell} -
q \ell)^2} \ H_n\big({x \over \ell} - q \ell\big),            \eqno(2.7)$$
where the
$H_n$
are Hermite polynomials,
and where
$ C_n \equiv \big( 2^{n+1} n! \pi^{3 \over 2} \ell\big)^{-{1 \over 2}} $.
These states are normalized according to
$ \int d^2r \ \phi^*_{nq} \phi_{n' q'} = \delta_{n n'}
\ \delta(q - q') $.

The Green's functions $G$ and $G^0$ may be related
by the Dyson equation
$$ G({\bf r},{\bf r}',s) = G^0({\bf r},{\bf r}',s)
+ \int d^2r'' \ G^0({\bf r},{\bf r}'',s) \ V({\bf r}'')
\ G({\bf r}'',{\bf r}',s) .                               \eqno(2.8)$$
For large $|{\bf r}-{\bf r}'|$, the magnitude of the Green's
function $G^0({\bf r},{\bf r}',s)$ falls off as a
Gaussian
$ e^{- {1 \over 4}|{\bf r}-{\bf r}'|^2 /\ell^2},$
except at its poles.
Then
$$ G({\bf r},{\bf r}',s) = G^0({\bf r},{\bf r}',s)$$
$$ + \  \int d^2r'' \ G^0({\bf r},{\bf r}'',s)
\big[ V({\bf \bar r}) + ({\bf r}'' - {\bf \bar r}) \cdot
\nabla V({\bf \bar r}) \big]
\ G({\bf r}'',{\bf r}',s) ,                               \eqno(2.9)$$
where ${\bf \bar r}$ is any point near ${\bf r}$, and where
higher order gradient terms have been neglected.
Equation (2.9) is now solved iteratively, keeping all terms containing
no gradients and all terms containing one
local gradient.
This leads to
$$ G({\bf r},{\bf r}',s) = G^1({\bf r},{\bf r}',s)
+ \int d^2r'' \ G^1({\bf r},{\bf r}'',s)
\ ({\bf r}'' - {\bf \bar r}) \cdot \nabla V({\bf \bar r})
\ G^1({\bf r}'',{\bf r}',s) ,                               \eqno(2.10)$$
where
$G^1({\bf r},{\bf r}',s)$
satisfies
$$ G^1({\bf r},{\bf r}',s) = G^0({\bf r},{\bf r}',s)
+ V({\bf \bar r})  \int d^2r'' \ G^0({\bf r},{\bf r}'',s)
\ G^1({\bf r}'',{\bf r}',s) .                               \eqno(2.11)$$
For notational simplicity, the dependence of $G^1$
on ${\bf \bar r}$ has been suppressed.
The solution of the integral
equation (2.11) is
$$ G^1\big({\bf r},{\bf r}',s\big)
 = G^0\big({\bf r},{\bf r}',s-V({\bf \bar r}) \big),      \eqno(2.12)$$
valid for any ${\bf \bar r}$ near ${\bf r}$.
The Green's function $G^0$, when renormalized by the local potential
$V({\bf \bar r})$, simply has its energy argument shifted
by $V({\bf \bar r})$.
Because of the arbitrariness in the choice of ${\bf \bar r}$,
$G^1$ is not unique. However, the effect of a change of
${\bf \bar r}$ on $G^1$ is compensated for by the corresponding change in the
 second
term in (2.10), and the complete Green's function (2.10) is independent
of ${\bf \bar r}$ to first order in the local gradient
$\nabla V({\bf \bar r})$.

At this point we have obtained a gradient expansion for
the Green's function $G$.
Unfortunately, the expression (2.10) contains all matrix elements
$\langle nq | {\bf r} | n' q' \rangle$
of ${\bf r}$ in the basis (2.7).
To circumvent this, we shall perform a gauge transformation,
for each ${\bf \bar r}$,
which rotates
the direction of the vector potential so that it is perpendicular
to the local gradient of the confining potential.
The gauge-transformed Green's functions may then be written in terms of
eigenfunctions which are localized in the
$\nabla V({\bf \bar r})$
direction, resulting in a simple closed-form expression for $G$.

To this end, we shall use the slowly varying function
$V({\bf r})$ to define a local orthonormal basis
${\bf n}_a \ \ (a=1,2)$
on the $z=0$ plane:
$$ {\bf n}_1({\bf r}) \equiv { \nabla V({\bf r}) \over
| \nabla V({\bf r})|}                       \eqno(2.13a)$$
$$ {\bf n}_2({\bf r}) \equiv {\bf e}_z \times
 {\bf n}_1({\bf r})   .                   \eqno(2.13b)$$
These basis vectors clearly satisfy the orthonormality
condition
$ {\bf n}_a \cdot {\bf n}_b = \delta_{ab}$
and are oriented according to
$ {\bf n}_1 \times {\bf n}_2 = {\bf e}_z $.
A vector potential directed parallel to
${\bf n}_2({\bf \bar r})$, for some fixed ${\bf \bar r}$,
and hence directed perpendicular to $\nabla V({\bf \bar r})$,
is given by
$$ {\bf A}' \equiv B \ [ {\bf n}_1({\bf \bar r}) \cdot {\bf r}]
\ {\bf n}_2({\bf \bar r}) .                         \eqno(2.14)$$
We suppress the
parametric dependence of ${\bf A}'$ on ${\bf \bar r}$.
This vector potential describes a uniform magnetic
field $(\nabla \times {\bf A}' = {\bf B})$ and is transverse
$(\nabla \cdot {\bf A}' = 0)$.
The normalized eigenstates of $H^0$ in this gauge
are
$$ \psi_{nq}({\bf r}) = C_n \ e^{-i q \eta \ell}
\ e^{-{1 \over 2}( \xi - q \ell)^2}
\ H_n(\xi - q \ell) ,          \eqno(2.15)$$
where
$ \xi \equiv {\bf n}_1({\bf \bar r}) \cdot {\bf r} / \ell $
and
$ \eta \equiv {\bf n}_2({\bf \bar r}) \cdot {\bf r} / \ell .$
The dependence of $\xi$ and $\eta$ on
${\bf \bar r}$ has also been suppressed.

The Green's function $G_{A'}$, computed in the gauge
${\bf A}' = {\bf A} + \nabla \Lambda$,
is related to
the Green's function $G_A$ in the original gauge by
$$ G_A({\bf r},{\bf r}',s) = e^{{i e \over \hbar c}[\Lambda({\bf r})
- \Lambda({\bf r}')]} \ G_{A'}({\bf r},{\bf r}',s) .        \eqno(2.16)$$
This allows one to compute a given matrix element
(${\bf r}$ and ${\bf r}'$ regarded as matrix indices)
of $G_A$ by transforming to some gauge ${\bf A}'$ where
$G_{A'}$ is simpler.
One may choose a different gauge for each matrix element.
The generator of the gauge transformation
from ${\bf A}=B x {\bf e}_y$
to ${\bf A}^{'}$ is given by
$$ \nabla \Lambda({\bf r}) = B \ [ {\bf n}_1({\bf \bar r}) \cdot {\bf r}]
\ {\bf n}_2({\bf \bar r}) - B x {\bf e}_y        .     \eqno(2.17)$$
Then we obtain
$$ G_A({\bf r},{\bf r}',s) = e^{{i e \over \hbar c}[\Lambda ({\bf r})
- \Lambda ({\bf r}')]}
\bigg[ \sum_n \int_{- \infty}^{\infty} dq
\ {\psi_{nq}({\bf r}) \psi^*_{nq}({\bf r}') \over
s - \hbar \omega_{\rm c}(n + \half) - V({\bf \bar r}) } $$
$$ + \ \nabla V({\bf \bar r}) \cdot
\sum_{n n'} \int_{-\infty}^{\infty} dq \ dq'
\ { [ \langle nq | {\bf r} | n' q' \rangle - {\bf \bar r} \ \delta_{nn'}
\ \delta (q - q') ] \ \psi_{nq}({\bf r}) \psi^*_{n' q'}({\bf r}')
\over
[s - \hbar \omega_{\rm c}(n + \half) - V({\bf \bar r})]
[s - \hbar \omega_{\rm c}(n' + \half) - V({\bf \bar r})] }
\bigg]                                       ,
 \eqno(2.18)$$
where $\Lambda$ is given by (2.17).
The matrix elements
$ \langle nq | {\bf r}|n' q' \rangle$
are now in the basis (2.15).
Because $\nabla V({\bf \bar r})$ points in the
${\bf n}_1({\bf \bar r})$ direction, only the matrix elements
of ${\bf n}_1 \cdot {\bf r}$ are required.
In what follows, we shall always choose ${\bf \bar r} = {\bf r}$;
the symmetric choice
${\bf \bar r} = \half ({\bf r}+{\bf r}')$ yields
identical results.
This concludes our construction of the Green's function for the 2D
electron gas in a magnetic field and a  slowly varying confining potential.

We now calculate the equilibrium density of the
confined electron gas.
The first term in (2.18) contributes an amount
$$ { 1 \over 2 \pi \ell^2} \sum_n \ f \big[\hbar \omega_{\rm c}
(n + {\half}) + V({\bf r}) - \mu \big].
 \eqno(2.19)$$
The contribution to the density from the second term in (2.18) vanishes.
Therefore, the final result is that stated
in (1.3).

We shall calculate the equilibrium current density by expanding
${\bf j}({\bf r})$ in the local basis ${\bf n}_a$,
$$ {\bf j}({\bf r}) = j_1({\bf r}) \ {\bf n}_1({\bf r})
+ j_2({\bf r}) \ {\bf n}_2({\bf r}),             \eqno(2.20)$$
where
$ j_a({\bf r}) \equiv {\bf j}({\bf r}) \cdot
{\bf n}_a({\bf r}).$
The component of the current density along the local potential
gradient is
$$ j_1 = - {e \over m} \oint {ds \over 2 \pi i} \ f[s-\mu]
\ \lim_{{\bf r}' \rightarrow {\bf r}} \ {\rm Re}
\bigg( - i \hbar \ {\bf n}_1 \cdot \nabla + {e \over c}
{\bf n}_1 \cdot {\bf A} \bigg) G({\bf r},{\bf r}',s) .       \eqno(2.21)$$
Using (2.18)
leads to
$$ j_1 = - e \omega_{\rm c} \ell^2 {e \over \hbar c}
\ {\bf n}_1 \cdot \nabla \Lambda \ \rho
\ - \ e \omega_{\rm c} x \ {\bf n}_1 \cdot {\bf e}_y \ \rho .  \eqno(2.22)$$
The first term in (2.22) comes from the ${\bf n}_1 \cdot \nabla$
acting on the exponential in (2.18), and the second term comes
from the diamagnetic part of (2.21).
There is no contribution from ${\bf n}_1 \cdot \nabla$ acting on
$G_{A'}$ (the quantity in square brackets) because these derivatives are real.
Using (2.17), we see that $j_1$ vanishes.
The transverse current density is given by
$$ j_2 = e \omega_{\rm c} \ell^2 \oint {ds \over 2 \pi i}
\ f[s-\mu]  \ {\rm Re}
\bigg( - {e \over \hbar c} {\bf n}_2 \cdot \nabla \Lambda ({\bf r}) \
 G_{A'}({\bf r},
{\bf r},s) $$
$$ + \ i \lim_{{\bf r}' \rightarrow {\bf r}} {\bf n}_2 \cdot
\nabla G_{A'}({\bf r},{\bf r}',s) \bigg)
- {e^2 \over mc} B x \ {\bf n}_2 \cdot {\bf e}_y \ \rho .         \eqno(2.23)$$
A straightforward calculation leads to
$$ j_2 =
 - {e^2 \omega_{\rm c} \ell^2 \over \hbar c} \ \rho
\ {\bf n}_2 \cdot \nabla \Lambda
\ + \ e \omega_{\rm c} \ell \ \xi \ \rho $$
$$ - \ e \omega_{\rm c} \ell^2 \sum_n (n+\half)
|\nabla \rho_n|
\ + \ e \omega_{\rm c} \ell^2  { |\nabla V| \over \hbar
\omega_{\rm c}} \ \rho
\ - \ e \omega_{\rm c} x \ \rho \ {\bf n}_2 \cdot {\bf e}_y,   \eqno(2.24)$$
where
$ \rho_n \equiv
 (1 / 2 \pi \ell^2)
\ f\big[\hbar\omega_{\rm c}(n+\half)
+ V({\bf r}) - \mu \big].$
Finally, after using (2.17), we find
$$ j_2 = - e \omega_{\rm c} \ell^2 \sum_n (n+\half)
|\nabla \rho_n|
\ + \ e \omega_{\rm c} \ell^2  {|\nabla V| \over \hbar
\omega_{\rm c}}  \ \rho            .                \eqno(2.25) $$
Note that $\nabla \rho$ is antiparallel to $\nabla V$.
Therefore, the equilibrium current density is given by the expression
stated in Section I.

\head{III. COMPARISON WITH AN EXACTLY SOLVABLE CASE}

In this section, we compare our results to the exact
equilibrium
density and current distributions
of a noninteracting 2D electron gas in a uniform magnetic
field ${\bf B}=B {\bf e}_z$
and a parabolic confining potential
$$ V(x) = {\half} m \omega_0^2 x^2  .         \eqno(3.1)$$
The results derived above apply to the case where
$\omega_0 < < \omega_{\rm c}$, so that the potential is
slowly varying over a range of several magnetic lengths.
In the gauge
${\bf A} = B x {\bf e}_y$,
the exact eigenstates and eigenvalues are
$$ \Psi_{nk} = (2^{n+1} n! \pi^{3\over 2} L)^{-{1 \over 2}}
\ e^{-iky}
\ e^{-{1 \over 2}({x \over L} - {\omega_{\rm c} \over \Omega} k L
)^2} \ H_n\big({x \over L} - {\omega_{\rm c}  \over \Omega} k L\big)
 \eqno(3.2)$$
$$ E_{nk} = \hbar \Omega (n + {\half}) + V(k L^2),      \eqno(3.3)$$
where
$\Omega^2 \equiv \omega_0^2 + \omega_{\rm c}^2$,
$L^2 \equiv \hbar / m \Omega$, and
where the $V$ appearing in (3.3) refers to (3.1). The states
$\Psi_{nk}$ are normalized as in Section II.

The exact equilibrium number density is given by
$$ \rho = \sum_n \int_{-\infty}^{\infty} dk
\ f[\hbar \Omega (n + {\half}) + V(k L^2) -\mu ] \ |\Psi_{nk}|^2,  \eqno(3.4)$$
where $\mu$ is the chemical potential.
Using (3.2), this may be
written as
$$ \rho = {1 \over 2 \pi \ell^2} \bigg( 1 +  {\omega_0^2 \over \omega_{\rm
c}^2}
 \bigg)
\sum_n \ \big(2^n n! \pi^{1 \over 2}\big)^{-1}
\int_{-\infty}^{\infty} dK \ f[ \hbar
\Omega (n+{\half}) +
{\Omega^2 \over \omega_{\rm c}^2} V(K L) - \mu ]$$
$$ \times \ e^{-({x \over L} - K)^2}
\ H_n^2\big({x \over L} - K \big),                        \eqno(3.5)$$
where
$K \equiv k L \Omega / \omega_{\rm c}$.
To facilitate comparison with our general expressions, we have written the
prefactor in terms of the magnetic length $\ell$ rather than $L$.

In Fig. 1, we compare the exact ground-state density
with the approximate distribution (1.3),
for the case $\mu = 3 \hbar \omega_{\rm c}$,
where there are three Landau levels filled
in the center of the well.
The curvature of the confining potential is chosen to be
$\omega_0 = {1 \over 20} \omega_{\rm c}$.
The principal difference between the exact density profile (dashed curve)
and the approximate profile (solid curve) is that the latter
neglects the detail at the step edges. The actual density
at the edges changes
over a few magnetic lengths, and in a manner which depends on the
particular Landau level involved.

There are two other differences
between the density profiles, which are too small to be visible in
Fig. 1. The first is that the density of a filled Landau level in the
parabolically confined system is slightly greater than
$(2 \pi \ell^2)^{-1}$, as is evident from the prefactor
of (3.5).
This reflects the small compressibility of the 2D electron gas at
integer filling factors, which appears as a response to the second derivative
of the confining potential, $V''(x),$
and which is neglected in (1.3).
The second difference is that the length scale
$L$ appearing in (3.5) is slightly less than the magnetic
length $\ell$,
$$ L = \bigg( 1 -
{\omega_0^2 \over 4 \omega_{\rm c}^2} \bigg) \ell . \eqno(3.6)$$
Hence, the actual density profile is slightly contracted relative to
the profile given by (1.3).

The exact equilibrium current density, which is directed in the $y$ direction,
may be written as
$$ j = \sum_n \int_{-\infty}^{\infty} dk
\ f[  \hbar
\Omega (n+{\half}) +
 V(k L^2) - \mu ] \ j_{nk} ,                       \eqno(3.7)$$
where
$$ j_{nk} \equiv - {e \over m} \ {\rm Re} \ \Psi^*_{nk} \big(
- i \hbar {\partial \over \partial y} + {e \over c}{\bf A}
\cdot {\bf e}_y \big) \Psi_{nk}                 \eqno(3.8)$$
is the contribution to the current density from the state
$\Psi_{nk}$.
 From (3.2), we obtain
$ j_{nk} = e \omega_{\rm c} \big( k \ell^2
- x \big) |\Psi_{nk}|^2,$
which may be rewritten as
$$ j_{nk} = { e L \Omega^2 \over \omega_{\rm c}} \big( K - {x \over L} \big)
\ |\Psi_{nk}|^2
+ e \omega_{\rm c} \ell^2 \bigg( { V'(x) \over \hbar \omega_{\rm c}}\bigg)
\ |\Psi_{nk}|^2       .                     \eqno(3.9)$$
Therefore, the current distribution may be written as
the sum of a bulk current and an edge current,
where the
bulk contribution is exactly
$$ j_{\rm bulk} = e \omega_{\rm c} \ell^2 \bigg({ V'(x) \over \hbar \omega_{\rm
 c}}\bigg)
\ \rho (x)                          ,                  \eqno(3.10)$$
and where
$$ j_{\rm edge} = {e \omega_{\rm c} \over 2 \pi \ell}
\bigg( 1 + {\omega_0^2 \over \omega_{\rm c}^2} \bigg)^{7 \over 4}
\sum_n
(2^n n! \pi^{1 \over 2})^{-1}
\int_{-\infty}^{\infty} dK \ f [ \hbar \Omega(n+{\half})
+ {\Omega^2 \over \omega_{\rm c}^2} V\big(K L\big) - \mu ]$$
$$ \times \ \big( K - {x \over L}\big)
\ e^{-(K - {x \over L})^2}
\ H_n^2\big( K - {x \over L}  \big) .                         \eqno(3.11)$$

In Fig. 2 we compare the exact ground-state current
distribution (dashed curve)
to the
current distribution given in Section I (solid curve).
The bulk contributions to the current density are nearly identical; they differ
only in that the density in (1.2) and the exact density in
(3.10) are slightly different, as shown in Fig. 1.
The sharp steps in $\rho$ lead to the sharp zig-zag
structure in the solid curve of Fig. 2.
However, we see that the approximate edge current
(1.1), which at zero temperature consists of a series of
$\delta$-functions, does not capture the form present in
the exact edge current (3.11).
However, as we shall show, the edge current (1.1) correctly accounts for the
{\it net} current associated with a given edge (the integrated edge current
density), in accordance with the earlier assertion that our
distributions correctly describe the large-scale features of the
exact distributions.

To prove this, we define the integrated edge currents,
$I_n^+$ and $I_n^-$, associated with the two edges of the
$n$th Landau level, one edge located to the right $(+)$ of
the origin and the other located to the left $(-)$.
$I_n^+$ is defined by
$$ I_n^+ \equiv
{e \omega_{\rm c} \over 2 \pi \ell} \bigg( 1
+ {\omega_0^2 \over \omega_{\rm c}^2} \bigg)^{7 \over 4}
(2^n n! \pi^{1 \over 2} )^{-1}
\int_0^{\infty} dx \int_{-\infty}^{\infty} dK
\ f[ \hbar \Omega(n+\half)
+ {\Omega^2 \over \omega_{\rm c}^2} V\big(K L\big) - \mu ]$$
$$ \times \ \big( K - {x \over L}\big)
\ e^{-(K - {x \over L})^2}
\ H_n^2\big( K - {x \over L}  \big).                           \eqno(3.12)$$
The definition of $I_n^-$, identical to (3.12) except that the $x$
integration is from $-\infty$ to $0$, shows that
$ I_n^- = - I_n^+ .$
This means that the integrated currents carried along the two
edges of each Landau level are of equal magnitude
and are opposite in sign, as is well-known.
In Appendix I,  we show that the
integrated ground-state edge currents are equal to
$$ I_n^{\pm} = \mp \big(n + {\half}\big)
{e \omega_{\rm c} \over 2 \pi} \bigg( 1 + {\omega_0^2 \over \omega_{\rm c}^2}
\bigg)^{3 \over 2}.                             \eqno(3.13)$$
This result is also valid for low temperatures such that
$k_{\rm \s B} T << \hbar \omega_{\rm c}$.
The integrated edge currents given by the
distribution (1.1) are easily shown to be
$$ I_n^{\pm} = \mp \big( n + {\half} \big)
{e \omega_{\rm c} \over 2 \pi},                  \eqno(3.14)$$
which is equal to (3.13), apart
from small corrections of order $\omega_0^2 / \omega_{\rm c}^2$.

Fig. 2 also demonstrates that the direction of the current oscillates with
 position.
This striking feature is correctly accounted for in our general expression by
noting that $\nabla \rho$ is antiparallel to $\nabla V$.
These oscillations, which originate from the oscillations in the magnetization
of the 2D electron gas as a function of filling factor, are a generic
feature of the current distribution in a confined 2D electron gas.
This is explained further in Appendix II.

\head{IV. APPLICATIONS}

\subhead{A. Current distribution in a  Hall bar}

The general expression we have derived for the low-temperature
current distribution in a 2D electron gas can be easily used with a
 self-consistent
potential obtained by solving the Hartree, Hartree-Fock, or
Kohn-Sham equations.
We now apply our result
to a stepped potential characteristic of the
low-temperature self-consistent Hartree potential of a narrow gate-confined
Hall bar.
The Hall bar is assumed to lie along the $y$ direction, with a confining
potential $V(x)$ and chemical potential as shown in Fig. 3.
A uniform magnetic field is applied in the $z$ direction.

The confining potential we use is similar to that obtained classically by
Chklovskii \etal\refto{6} for a narrow gate-confined
channel, and confirmed by self-consistent Hartree and Hartree-Fock
calculations.\refto{9-12}.
However, we have approximated the potential by a series
of linear potentials and we have included small slopes
$(10^{-3} \ \hbar \omega_{\rm c} / \ell)$,
on the plateaus and in the central region, to include, in a simple fashion,
the effects of imperfect screening.
These slopes are too small to be resolved in Fig 3.
We shall show that this piecewise linear potential, which supports
a low-temperature density distribution consisting of wide compressible
edge channels and narrow incompressible strips, is sufficient
to accurately characterize the
low-temperature current distribution in a narrow Hall bar.

In Fig. 4, we plot the low-temperature
$( k_{\rm \s B} T = 0.002 \ \hbar \omega_{\rm c})$
density and current distributions in the
confining potential of Fig. 3.
The large number of fractionally occupied states in
the compressible edge regions leads to
smooth edge profiles (solid curve).
The edge channels occur in the plateaus of the potential, whereas the
incompressible strips occur at the potential steps.
The current distribution (dashed curve) consists of edge currents in the
compressible regions and bulk currents in the incompressible regions.
Small steps occur in the density (at $x=\pm 15\ell$) and
current density (at $x=0$)
because of the sharp corners in our piecewise linear potential.

\subhead{B. Universal integrated currents}

In the ideal case of perfect compressibility and incompressibility,
only one type of current contributes to a given region. We now show
that, in this ideal case, the integrated currents in these
regions are universal, independent of the size and position of the regions, and
 the
geometry of the sample.
It is easy to calculate total current carried in each edge channel and
in each incompressible strip (we neglect
incompressible strips at fractional filling factors).
{}From (1.1), we see that the integrated edge current depends only on the
net density change across the edge channel, which is $(2 \pi \ell^2)^{-1}$
at low temperatures, and on the index $n$ of the Landau states which form
the edge. Therefore, an edge channel at the edge of the $n$th Landau
level carries a current of magnitude
$$ I_{{\rm edge},n} = (n + \half) {e \omega_{\rm c} \over 2 \pi} ,
 \eqno(4.1)$$
independent of the width  and position of the edge channel.
A central compressible  region  which supports no net density change across its
boundaries, carries no integrated current.
{}From (1.2), we see that the integrated
bulk current in an incompressible strip with integral filling factor $\nu$ is
simply
$$ I_{\rm bulk} = \nu {e \omega_{\rm c} \over 2 \pi} ,         \eqno(4.2)$$
independent of the width  and position of the strip.
Fig. 4 also demonstrates the alternating pattern
of the directions of the edge and bulk currents.

In nonideal cases, for example at higher temperature,
both the edge and bulk components
contribute simultaneously, and there are corrections to (4.1) and (4.2).
However, the oscillations in the directions of the currents generally
remains.

\subhead{C. Quantized persistent currents}

As a final application of our result, we shall investigate the quantization
of the persistent current (total azimuthal current through a radial cross
section)
in a quantum dot predicted recently by Avishai and Kohmoto$^{24}$.  Consider a
system of noninteracting electrons in a slowly varying cylindrically symmetric
potential $V(r)$ subjected to a uniform magnetic field ${\bf B} = B {\bf e}_z$.
The quantum dot is assumed to be large enough so that there are many degenerate
Landau  states in the bulk. The ground-state radial density and current
distributions will be similar to those in Figs. 1 and 2 (with $x$ acting  as a
radial coordinate), except that the central incompressible region will be
larger
and the bulk currents will vanish there because of the assumed flatness of the
confining potential.  Following Ref. 24, we shall calculate the integrated
azimuthal current
$$ I=  \int_0^\infty  dr \ j(r),
 \eqno(4.3)$$
when the Fermi energy lies in a bulk Landau level.  We shall initially assume,
for
simplicity, that the Fermi energy is just below Landau level $n$ so that the
filling factor is $\nu = n$ in the center of the dot.  Afterwards, we treat the
realistic situation  where the Fermi energy is somewhere in the bulk states of
Landau level $n$.

We shall evaluate the persistent current by dividing the integral (4.3) into
$n$
regions of filling factor $n, n-1,...,1$.  The central region is from $r=0$
(where
the azimuthal current density vanishes) to $r=r'$ where $V(r')-V(0) = \hbar
\omega_c$, and has constant filling factor $\nu=n$.  The integrated bulk
current in
this region is simply $ne \omega_c/2 \pi$.   The edge current concentrated
about
$r=r'$ comes from states with Landau level index $n-1$ and contributes an
amount
$-(n-\half)e \omega_c/2\pi$ to (4.3).  The integrated current in this first
region
is therefore $e \omega_c/4\pi$, independent of $n$.  Beyond $r=r'$, the filling
factor decreases to $\nu = n-1,n-2,...,0$.  It is simple to verify that the
contribution to (4.3) from each of these regions is $e \omega_c/4\pi$.
Therefore,
whenever the Fermi energy lies just below the bulk states of Landau level $n$,
the
persistent current is $ne \omega_c/4\pi$.

As the number of electrons in the quantum dot is changed, the Fermi energy is
generally pinned in a set of bulk states, not below them.  Because the bulk
states
carry no bulk current, the persistent current calculated above is modified by
the
integrated edge current $-(n+\half) e \omega_c/2\pi$ of Landau level $n$ only.
Therefore, when the Fermi energy is locked in the bulk states of Landau level
$n$,
the persistent current (4.3) is
$$ I=  - (n+1) {e \omega_c \over 4 \pi}.
 \eqno(4.4)$$
Avishai and Kohmoto$^{24}$ have also predicted a persitent current which is
quantized in integer multiples of $e \omega_c/4\pi$ as a function of particle
number.

\head{ACKNOWLEDGMENTS}

This work was supported by the National Science Foundation
through Grant No. DMR-9100988.
We thank
Rolf Gerhardts,
Yigal Meir,
Boris Shklovskii,
and David Thouless,
for providing us with preprints of their work.

\head{APPENDIX I}

Here we calculate the
integrated ground-state edge currents, $I_n^+,$ which may be written as
$$ I_n^+ \equiv g \ A_n \int_0^{\infty} dx \int_{-\infty}^{\infty} dK
\ \Theta [ \mu - \hbar \Omega(n+{1 \over 2})
- {\Omega^2 \over \omega_{\rm c}^2} V\big(K L\big)  ]$$
$$ \times \ \big( K - {x \over L}\big)
\ e^{-(K - {x \over L})^2}
\ H_n^2\big( K - {x \over L}  \big),                           \eqno(A1)$$
where
$A_n \equiv (2^n n! \pi^{1 \over 2} )^{-1},$
$ g \equiv (e \omega_{\rm c} / 2 \pi \ell) [ 1
+ (\omega_0 / \omega_{\rm c})^2]^{7 \over 4},$
and where $\Theta[\epsilon]$ is the unit step function.
The Hermite polynomial relations
$x \ H_n = {\half} \ H_{n+1} + n \ H_{n-1}$
and
$H_n' = 2 n H_{n-1}$
lead to
$$ I_n^+ = - {\half} g L \ A_n \int_0^{\infty} dX \int_{-K_n}^{K_n} dK
\ e^{-(X - K)^2} $$
$$ \times \ H_n(X - K) \bigg[ H_{n+1}(X - K) +
{ \partial \over \partial X} H_n(X - K) \bigg],                     \eqno(A2)$$
where
$X \equiv x/L$, and where
$K_n L$, defined by
$ \mu =  \hbar \Omega(n + {1 \over 2}) + (\Omega^2 / \omega_{\rm c}^2)
V(K_n L) $,
is the center of the $n$th edge.
By using
$H_n = \half H_{n+1}'/(n+1)$,
we find
$$ I_n^+ = - {\textstyle {1 \over 4}} g L \ A_n
\int_0^{\infty} dX \int_{-K_n}^{K_n} dK
\ e^{-(X - K)^2} $$
$$ \times \  {\partial \over \partial X} \bigg(
{ H_{n+1}^2(X - K) \over 2 (n+1)} + H_n^2(X-K) \bigg)  .     \eqno(A3)$$
After integrating by parts, assuming that
$K_n >> 1$ (which is equivalent
to $\omega_0 < < \omega_{\rm c}$),
we obtain the recurrence relation
$ I_{n+1}^+ = I_n^+ - g L$,
which leads to
$ I_n^+ = I_0^+ - n g L  .$
{}From (A1), we find
$I_0^+ = - {1 \over 2} g L$.
Therefore, the zero-temperature integrated edge currents are equal to
$$ I_n^{\pm} = \mp \big(n + {\half}\big)
{e \omega_{\rm c} \over 2 \pi} \bigg( 1 + {\omega_0^2 \over \omega_{\rm c}^2}
\bigg)^{3 \over 2},                                           \eqno(A4)$$
as stated.

\head{APPENDIX II}

In this appendix, we shall derive the current distribution
given in Section I
from linear response theory
to provide a deeper understanding of the
result and to establish the connection with
the distribution derived from
current-density functional theory.\refto{1}
We start with a uniform electron gas in a magnetic
field ${\bf B} = B {\bf e}_z$ at low
temperature $(k_{\rm \s B} T < < \hbar \omega_{\rm c})$.
The static
current-current response function
$\chi^{\mu \nu}({\bf q}) \  (\mu, \nu = 0,1,2) $, defined by
$ j^\mu({\bf q}) = \chi^{\mu \nu}({\bf q}) \ A^\nu({\bf q}),$
describes the Fourier components of the three-current
$j^\mu \equiv(\rho,{\bf j})$ induced by the application
of an infinitesimal potential
$A^\mu \equiv (V,{\bf A})$.
Combining
$\chi^{00}({\bf q})$ and
$\chi^{i0}({\bf q})$
together
leads to the nonlocal relation
$$ j^i({\bf q}) = { \chi^{i0}({\bf q}) \over \chi^{00}({\bf q})}
\ \rho({\bf q})                                   \eqno(B1)$$
between current and density.
It is simple to show that
$\chi^{i0} = - i c ( {\bf q} \times {\bf e}_z )^i \big( \partial M
/ \partial \mu \big)_{B}$
and
$\chi^{00} = - \big( \partial \rho
/ \partial \mu \big)_{B}$
in the long wavelength limit,
where $M$ is the orbital magnetization of a
uniform 2D electron gas.
The Maxwell relation
$\big( \partial M / \partial \mu \big)_B
= \big( \partial \rho / \partial B \big)_{\mu}$
then leads to
$ {\bf j}({\bf r}) = - (e \hbar / 2 m)
\ \gamma \ \nabla \rho ({\bf r}) \times {\bf e}_z ,$
where
$$ \gamma \equiv - {2 m c \over e \hbar}
\ { \big(\partial \rho / \partial B \big)_{\mu}
\over \big(\partial \rho / \partial \mu \big)_B }  .   \eqno(B2)$$
This is equivalent to the LDA result derived in Ref. 1.

In the uniform electron gas,
$$ \bigg( {\partial \rho \over \partial B} \bigg)_{\mu}
= {\rho \over B} - {1 \over 2 \pi \ell^2} \bigg({e \hbar \over mc}\bigg) \sum_n
(n + \half) \ F^{-1}_n(\mu)                  \eqno(B3)$$
and
$$ \bigg( {\partial \rho \over \partial \mu} \bigg)_{B}
=  {1 \over 2 \pi \ell^2} \sum_n \ F^{-1}_n(\mu),         \eqno(B4)$$
where
$ F_n(\mu) \equiv  4 k_{\rm \s B} T
\ {\rm cosh}^2 \big( [\mu - \hbar \omega_{\rm c}(n+\half)]
/ 2 k_{\rm \s B} T \big) $.
At low temperatures,
$F_n^{-1}(\mu)$ is strongly localized about
$\mu = \hbar \omega_{\rm c} (n+\half)$
with a width of order
$k_{\rm \s B} T$.
Note that the denominator in (B2) is positive definite, whereas
the numerator is a
sum of two terms with opposite signs.
We shall show that the first term in (B3) largely determines the current
density
in the incompressible regions while the second term largely determines the
current
in the compressible regions.

In a compressible region  near filling factor $\nu = n + \half$,
the chemical potential is
$\mu \approx \hbar \omega_{\rm c}(n+\half)$.
The narrow range
$|\mu - \hbar \omega_{\rm c} | \leq k_{\rm \s B} T$
corresponds to the entire range of compressible densities.
Because only a single term contributes to the summations (apart from
terms exponentially small in $k_{\rm \s B} T / \hbar \omega_{\rm c}$),
$$ \gamma = - { 2 \rho \over \hbar \omega_{\rm c} }
\bigg( {\partial \rho \over \partial \mu } \bigg)_B^{-1}
\ + \ (2n+1) .                           \eqno(B5)$$
Then
$ \nabla \rho = - \big( \partial \rho / \partial \mu \big)_B \ \nabla V$
leads to the distribution given in (1.1) and (1.2),
at filling factor
$\nu = n + \half$.
The current density in a compressible region, where the self-consistent
potential is nearly uniform, is largely determined by the second term in (B5).
In an incompressible region, for example  near $\nu = n$,
the $F_n$ terms vanish with exponential accuracy, and
 the bulk term (1.2)  dominates.

By construction, linear response theory is valid only when the
perturbing stimulus is infinitesimal. However, the results obtained in this
appendix
are apparently valid even when the applied potential
$V({\bf r})$ is large, as long as
$V({\bf r})$ is slowly varying so that the
long wavelength
limit applies.
We speculate that the reason for this is that the response
function
$\chi^{\mu \nu}({\bf r},{\bf r}')$,
a correlation function of the
three-currents $j^\mu({\bf r})$
and $j^\nu({\bf r}')$,
is short-ranged in $|{\bf r}-{\bf r}'|$.
Then, as we have demonstrated with the single-particle
Green's function, the effect of the confining potential
may be treated locally as a shift in the chemical potential,
plus the linear response of a uniform electron gas with this shifted
chemical potential to a  weak electric field.

\references

\refto{1} G. Vignale and P. Skudlarski, Phys. Rev. B {\bf 46}, 10232 (1992).
See also G. Vignale, in {\it Proceedings of the NATO ASI on
Density Functional Theory}, edited by E. K. U. Gross and R. M. Dreizler (in
 press).

\refto{2} G. Vignale and M. Rasolt, Phys. Rev. Lett. {\bf 59}, 2360 (1987);
 Phys. Rev.
B {\bf 37}, 10685 (1988).

\refto{3} C. W. J. Beenakker and H. van Houten, in {\it Solid State Physics}
 Vol. {\bf 44},
edited by H. Ehrenreich and D. Turnbull (Academic Press, New York, 1991).

\refto{4} M. B\"uttiker, in {\it Semiconductors and Semimetals} Vol. {\bf 35},
 edited
by M. Reed (Academic Press, New York, 1992).

\refto{5} D. B. Chklovskii, B. I. Shklovskii, and L. I. Glazman, Phys. Rev. B
 {\bf 46}, 4026 (1992).

\refto{6} P. L. McEuen \etal, Phys. Rev B {\bf 45}, 11419 (1992).

\refto{7} D. B. Chklovskii, K. A. Matveev, and B. I. Shklovskii, Phys. Rev. B
 {\bf 47}, 12605 (1993).

\refto{8} M. M. Fogler, E. I. Levin, and B. I. Shklovskii (to be published).

\refto{9} B. Y. Gelfand and B. I. Halperin, Phys. Rev. B {\bf 49}, 1862 (1994).

\refto{10} L. Brey, J. J. Palacios, and C. Tejedor, Phys. Rev. B {\bf 47},
13884
 (1993).

\refto{11} C. Wexler and D. J. Thouless, Phys. Rev. B {\bf 49}, 4815 (1994).

\refto{12} K. Lier and R. R. Gerhardts, (to be published).

\refto{13} C. de C. Chamon and X. G. Wen, Phys. Rev. B {\bf 49}, 8227 (1994).

\refto{14} Y. Meir, Phys. Rev. Lett. {\bf 72}, 2624 (1994).

\refto{15} A. H. MacDonald, T. M. Rice, and W. F. Brinkman, Phys. Rev. B {\bf
 28}, 3648 (1983).

\refto{16} O. Heinonen and P. L. Taylor, Phys. Rev. B {\bf 32}, 633 (1985).

\refto{17} D. J. Thouless, J. Phys. C {\bf 18}, 6211 (1985).

\refto{18} Q. Li and D. J. Thouless, Phys. Rev. Lett. {\bf 65}, 767 (1990).

\refto{19} D. Pfannkuche and J. Hajdu, Phys. Rev. B {\bf 46}, 7032 (1992).

\refto{20} D. J. Thouless, Phys. Rev. Lett. {\bf 71}, 1879 (1993).

\refto{21} O. Heinonen and M. D. Johnson, Phys. Rev Lett. {\bf 71}, 1447
(1993).

\refto{22} T. K. Ng, Phys. Rev. Lett. {\bf 68}, 1018 (1992).

\refto{23} S. Hershfield, Phys. Rev. Lett. {\bf 70}, 2134 (1993).

\refto{24} Y. Avishai and M. Kohmoto, Phys. Rev. Lett. {\bf 71},
279 (1993).

\endpage

\head{FIGURE CAPTIONS}

\vskip 0.50in

FIG. 1.$\ $ Ground-state density, $\rho(x)$,
in a parabolic potential with
curvature $\omega_0 = {1 \over 20} \omega_{\rm c}$ and chemical
potential $\mu = 3 \hbar \omega_{\rm c}$,
plotted in units of $\rho_0 \equiv (2 \pi \ell^2)^{-1}$.
The solid curve follows from
the expression
(1.3) in the text; the dashed curve is exact.
Lengths are plotted in units of magnetic length $\ell$.

\vskip 0.50in

FIG. 2.$\ $ Exact and approximate
$y$ components of the ground-state current density, $j(x)$,
in a parabolic potential, with the same parameters
as in Fig. 1. The current distributions are plotted in units
of $j_0 \equiv e \omega_{\rm c} / 2 \pi \ell$.
The solid curve follows from the distribution of Section I,
with the vertical arrows denoting $\delta$-functions.
The dashed curve is exact.
Lengths are plotted in units of magnetic length $\ell$.
Note the alternating directions, or signs, of the edge and bulk currents.

\vskip 0.50in

FIG. 3.$\ $ Piecewise linear confining potential (solid curve) characteristic
of the low-temperature self-consistent Hartree potential of a narrow
gate-confined Hall bar.
The potential is plotted in units of $\hbar \omega_{\rm c}$,
and the dashed line is
the chemical potential $\mu = {\textstyle {3 \over 2}} \hbar
\omega_{\rm c}$.
Lengths are plotted in units of magnetic length $\ell$.

\vskip 0.50in

FIG. 4.$\ $ Equilibrium density (solid curve) and current density (dashed
curve) corresponding to the stepped confining potential and
chemical potential of Fig. 3, at low temperature $(k_{\rm \s B} T
= 0.002 \ \hbar \omega_{\rm c})$. The density is plotted in units of
$\rho_0 \equiv (2 \pi \ell^2)^{-1}$ and the current density is plotted
in units of $j_0 \equiv e \omega_{\rm c} / 2 \pi \ell$.
Lengths are plotted in units of magnetic length $\ell$.
Note the alternating directions, or signs, of the edge and
bulk currents.

\endit